\documentclass[final,epj]{svjour}
\usepackage{hyperref}
\usepackage{amsmath}
\usepackage{bm}
\usepackage{graphicx}
\usepackage{latexsym}

\usepackage{setspace}

\usepackage{amssymb,color}

\begin{document}
\title{Synchronizability of chaotic logistic maps in delayed complex networks}

\titlerunning{Synchronizability of chaotic logistic maps...}

\author{ Marcelo Ponce C. \inst{1},
C. Masoller\inst{2}\and Arturo C. Mart\'{\i}\inst{1}}

\institute{Instituto de F\'{\i}sica, Facultad de   Ciencias, Universidad de la Rep\'ublica, Igu\'a 4225, Montevideo
  11400, Uruguay\and Departament de Fisica i Enginyeria   Nuclear, Universitat Politecnica de Catalunya, Colom 11, E-08222
  Terrassa, Barcelona, Spain}

\date{\today}
\abstract{
We study a network of coupled logistic maps whose interactions
occur with a certain distribution of delay times. The local dynamics
is chaotic in the absence of coupling and thus the network is
a paradigm of a complex system.
There are two regimes of synchronization, depending on the
distribution of delays: when the delays are sufficiently heterogeneous
the network synchronizes on a steady-state
(that is unstable for the uncoupled maps); when the delays are homogeneous,
it synchronizes in a time-dependent state (that is either periodic or chaotic).
Using two global indicators we quantify the synchronizability on the two regimes,
focusing on the roles of the network connectivity and the topology.
The connectivity is measured in terms of the
average number of links per node, and we consider various
topologies (scale-free, small-world, star, and nearest-neighbor with
and without a central hub). With weak connectivity and weak
coupling strength, the network displays an irregular oscillatory
dynamics that is largely independent of the topology and of the delay
distribution. With heterogeneous delays, we find a threshold connectivity level
below which the network does not synchronize, regardless of the network size.
This minimum average number of neighbors seems to be independent of the delay distribution.
We also analyze the effect of self-feedback loops and find
that they have an impact on the synchronizability of small networks
with large coupling strengths. The influence of feedback, enhancing
or degrading synchronization, depends on the topology and on the
distribution of delays.}


\PACS{05.45.Xt, 05.45.-a, 05.45.Ra, 05.45.Pq}

\maketitle

\section{Introduction}
Complex networks arise due to self-organization phenomena in many
real systems, such as food webs, the Internet, social networks,
genes, cells and neurons \cite{strogatz_nature_2001}. A lot of
research has been devoted to understanding the collective behavior
emerging in complex networks \cite{pramana}, given the individual dynamics of the
nodes and the coupling architecture. Ecological webs, for example,
describe species by means of nodes connected by links, representing
the interactions. The interactions can be either direct or indirect
through intermediate species, and can be of antagonistic type, such
as predation, parasitism, etc., or mutually beneficial, such as
those involving the pollination of flowers by insects \cite{sole}. A
study of the network architecture decompose food webs in spanning
trees and loop-forming links, revealing common principles underlying
the organization of different ecosystems \cite{food_webs}. Cells use
metabolic networks of interacting molecular components in processes
that generate mass, energy, information transfer and cell-fate
specification \cite{metabolic}. Adaptation and robustness have been
shown to be consequences of the network's connectivity and do not
require the 'fine-tuning' of parameters \cite{barkai}. In a cat's
brain, functional connectivity  has been studied within the
framework of complex networks \cite{kurths_brain}. In the human
brain, magnetic resonance imaging has been used to extract
functional networks connecting correlated human brain sites
\cite{victor_brain}. Analysis of the resulting networks in various
tasks (e.g., move a finger) has shown interesting features in the
brain, such as scale-free structure, a high clustering coefficient,
and a small characteristic path length.

Complex network are relevant not only from an academical point of
view but also from an applied perspective. Models based on complex
networks for the spread of diseases have identified mitigation
strategies for epidemic spread. In \cite{kuperman} the spread of an
infection was analyzed for different population structures, ranging
from ordered lattices to random graphs, and it was shown that for
the more ordered structures, there was a fluctuating endemic state
of low infection. In \cite{epidemic} it was shown that outbreaks can
be contained by a strategy of targeted vaccination combined with
early detection, thus avoiding mass vaccination of the hole
population. In communication networks, error tolerance and attack
vulnerability are key issues. In \cite{ataques} it was shown that
error tolerance is not shared by all networks that contain redundant
wiring, but is displayed only by those with scale-free topology.

The structure of a network is a key issue in determining its
functional properties. In real networks communities, or modules,
associated with highly interconnected parts, have been identified
\cite{communities,communities2}. A nice example of a community
structure has been unveil in networks of musical tastes
\cite{javier}, having also practical applications for the
development of commercial music recommendation engines.

Well-characterized modules have been identified in synthetic gene
networks \cite{genes}, where positive feedback and noise play
important roles for the repression and the activation of gene
expression. An excitable module containing positive and negative
feedback loops has been identified as a key mechanism inducing
transient cellular differentiation \cite{jordi}. A method for
classifying nodes into universal roles according to their intra- and
inter-module connections has been recently proposed \cite{modules}
and applied to the study metabolic networks.

The synchronizability of a network, or is propensity for
synchronization, is another key issue in determining the network
functional properties \cite{damian}. Why some topologies are easier
to synchronize than others, is still an issue not fully understood.
Heterogeneity in the connection strengths tends to enhance
synchronization \cite{motter05}. A weighting procedure based upon
the global structure of network pathways has been shown to improve
synchronizability \cite{sincronizability_stefano}. Dynamical
adaptation, where the coupling strengths develop according to the
local synchronization of the node and its neighbors, resulting in
weighted coupling strengths that are correlated with the topology,
also enhances synchronizability \cite{sincronizability_kurths}.
However, heterogeneity in the connectivity distribution can have the
opposite effect: networks with homogeneous connectivity have been
found to have larger propensity for synchronization than the
heterogeneous ones \cite{sincronizability_japones}.

The speed of transmission of information among the network also
affects the synchronizability. Instantaneous interactions have been
studied a lot in spite of the fact in many situations they are not
realistic, because the information propagates with a finite speed. A
more realistic scenario considers that the links have associated
delay times, and that the delays are the same for all the links. It
has been shown that the presence of such uniform delays in the
communications among the nodes can result in enhanced
synchronizability. In \cite{atay_PRL_2004} it was shown that, in a
network of chaotic maps, it was possible to synchronize the delayed
network where the undelayed network, with instantaneous links, did
not.

An even more realistic approach for complex and disordered systems
is to consider heterogeneously distributed delays. In previous work
\cite{nosotros_prl_2005,marti2006} we studied networks of coupled
maps with links that have heterogenous delays, and investigated the
relation of the network topology with its ability to synchronize. We
found that (i) the synchronizability was enhanced by random delays
as compared to networks with uniform delays, (ii) the network
synchronized in a steady state in the presence of random delays (in
contrast, with uniform delays the synchronization is in time a
dependent state, \cite{atay_PRL_2004}) and (iii) the
synchronizability depends mainly on the mean connectivity and is
rather independent of the topology.

The aim of this paper is to further analyze how the synchronizability
depends on the connectivity and on the topology, when there are delays
in the links among the nodes. We consider both regular and random
network topologies, covering the cases of homogeneous and
heterogeneous distribution of the links. We also study the effects of
centrality and locality on the dynamics of the array.  We characterize
the synchronizability in terms of two indicators, one that tends to
zero when the networks synchronizes, regardless if the synchronization
is in a time dependent or in a steady state, and the other that tends
to zero only when the synchronization is in a steady state. Using
these indicators we also analyze the impact of self-feedback
links. This paper is organized as follows.  Section~\ref{model}
presents the model, the different distributions of delays, and several
topologies used.  Section~\ref{resultados} presents the results of the
simulations, and, finally, Sec.~\ref{summary} presents a summary and
the conclusions.

\section{Model}
\label{model}

We consider $N$ logistic maps coupled as:
\begin{equation}
\label{eq:mapa} x_i(t+1)= (1-\epsilon) f[x_i(t)] + \frac{\epsilon}
{b_i} \sum_{j=1}^N \eta_{ij} f[x_j(t-\tau_{ij})],
\end{equation}
where $t$ is a discrete time index, $i$ is a discrete spatial index
($i=1\dots N$), $f(x)=ax(1-x)$ is the logistic map, $\epsilon$ is
the coupling strength and $\tau_{ij} \ge 0$ is the delay time in the
interaction between the $i$th and $j$th nodes (the delay times
$\tau_{ij}$ and $\tau_{ji}$ need not be equal). The matrix
$\eta=(\eta_{ij})$ defines the connectivity of the network:
$\eta_{ij}=\eta_{ji}=1$ if there is a link between the $i$th and
$j$th nodes, and zero otherwise. The sum in Eq.~(\ref{eq:mapa}) runs
over the $b_i$ nodes which are coupled to the $i$th node, $b_i =
\sum_j \eta_{ij}$. The normalized pre-factor $1/b_i$ means that each
map receives the same total input from its neighbors.

A particularly simple solution of Eq. (\ref{eq:mapa}) is such that
all the maps of the network are in a fixed point of the uncoupled
map, i.e.,
\begin{equation}
\label{eq:solution} x_i(t) = x_0 \;\;\;\; \forall \;\;\;\;i,
\end{equation}
with $x_0=f(x_0)$. While this solution exists for all delay
distributions ($\tau_{ij}$) and for all coupling topologies
($\eta_{ij}$), the statistical linear stability analysis performed
in \cite{nosotros_prl_2005} showed that this solution is unstable
unless the distribution of delays is wide enough. In the other
limiting case of all-equal delays, $\tau_{ij}=\tau$ $\forall$ $i$
and $j$, it was shown in Ref.\cite{atay_PRL_2004} that the network
synchronizes isochronally, in a spatially homogeneous time-dependent
state:
\begin{equation}
\label{eq:solution2} x_i(t) = x(t) \;\;\;\; \forall \;\;\;\;i,
\end{equation}
with $x(t+1)= (1-\epsilon) f[x(t)] + \epsilon f[x(t-\tau)]$.

\subsection{Distribution of delays}

We consider delays distributed as: $\tau_{ij} = \tau_0 +
\mathrm{near} (c \xi)$, where $c$ is a parameter that allows varying
the width of the delay distribution; $\xi$ is Gaussian distributed
with zero mean and standard deviation one; $\mathrm{near}$ denotes
the nearest integer. Depending on $\tau_0$ and $c$ the distribution
of delays is truncated to avoid negative delays.

The synchonizability of the network depends on both, the mean delay
and the width of the delay distribution \cite{nosotros_prl_2005}.
The larger the value of $\tau_0$, the larger the dispersion of the
delays has to be, for the network to synchronize in the steady state
(see Fig. 3 of \cite{nosotros_prl_2005}). A similar observation was
recently reported in \cite{japoneses_2008}, for an
integro-differential equation describing the collective dynamics of
a neural network with distributed signal delays. An interesting
interpretation of both observations is provided by the work of
Morgado et al. \cite{brazil_EPL_2007}: a certain degree of
randomness in a network (due to random delays, random connectivity,
or even random initial conditions) results in additive and/or
multiplicative noise terms in an ''effective'' single-node equation
of motion for $x_i(t+1)$, expressed in terms of $x_i(t)$ and
nonlinear feedback memory terms, that for the Logistic map are of
polynomial type in $x_i(t-n)$ with $1<n<t$. In
\cite{brazil_EPL_2007} it was found that the right memory profile
can create optimal conditions for synchronization, in other words,
an optimal memory range enhances a network propensity for
synchronization.

Since our aim is to study the roles of the topology and of the
connectivity, we keep fixed the distribution of delays, defined by
the parameters $\tau_0$ and $c$. However, we compare the results
obtained for distributed delays ($c\ne 0$) with those obtained for
all-equal delays ($c=0$) and with those obtained for no delays
($\tau_0=0$, $c=0$).

\subsection{Network connectivity and topology}

The connectivity of a network is measured in terms of the average
number of links per node, \begin{equation}\langle b
\rangle=\frac{1}{N}\sum_{i=1}^N b_i.\end{equation}

We consider the five topologies,
three of them are regular networks where the links are distributed
deterministically among the nodes, while the other two are
heterogenous networks where the links are distributed
stochastically, with given rules. The networks are:

(i) a nearest-neighbor network (referred to as NN network) with
periodic boundary conditions where each node $i$ is linked to its
neighboring nodes $i \pm 1, i \pm 2, . . . , i \pm K$, with $K$ an
integer. The number of neighbors is the same for all the nodes, and
$\langle b \rangle=2K$.

(ii) the NN network with the addition of a central node connected to
all other nodes (referred to as ST network). In this case, $N-1$
nodes have $2K+1$ links and one node has $N-1$ links. Thus, $\langle
b \rangle=[(2K+1)(N-1)+(N-1)]/N=(2K+2)(N-1)/N$.

(iii) a star-type network where there are $K$ central nodes that are
connected to all other nodes and $N-K$ peripheric nodes that are
connected only to the central ones (referred to as KA). In this
case, $K$ nodes have $N-1$ links and $N-K$ nodes have $K$ links.
Thus, $\langle b \rangle=[K(N-1)+(N-K)K]/N$. This network has a pure
star-type structure and is {\it centrally organized}, in contrast to
NN, that is {\it locally organized}.

(iv) a small-world (SW) network constructed according to the Newman
and Watts algorithm \cite{nw_1999}.

(v) a scale-free (SF) network constructed according to the Barabasi
and Albert method \cite{barabasi-RMP}.

Recently, the influence of feedback loops on the
synchronization of complex systems has received attention.
In networks of globally coupled units (self-sustained oscillators or maps) delayed feedback in the mean field can, depending on the feedback strength and on the delay time, enhance or suppress synchronization \cite{misha}.
In a system composed by
two phase oscillators with instantaneous mutual coupling, a
delayed feedback loop in each oscillator
enhances synchronizability if the coupling strength is not too strong, while degrades synchronizability if the coupling exceeds a threshold
value \cite{tass}. In small networks of mutually delayed-coupled
oscillators, which typically do not exhibit stable isochronal
synchronization, by including delayed feedback loops to the nodes, the
oscillators become isochronally synchronized \cite{kanter,Schwartz}.
To analyze the relevance of feedback loops we
consider two situations: the diagonal elements of the coupling
matrix, $\eta_{ii}$, are all set equal to $1$, or are all set equal
to $0$. In this way, in each node a feedback loop is included,
or is forbidden. The delays of the feedback loops, $\tau_{ii}$, have the same distribution as the delays of the
mutual interactions, $\tau_{ij}$.

\graphicspath{{c:/fortran/mapas/weak_coupling/}}


\subsection{Synchronization indicators}

To characterize the degree of synchronization and to distinguish
between steady state and time dependent synchronization, we use the
following indicators,
\begin{eqnarray}
\label{sp1}
 \sigma^2 &=& \frac{1}{N} \langle \sum_i (x_i- \langle x \rangle_s)^2
 \rangle_t \\
\label{sp2}
 \sigma'^2 &=&  \frac{1}{N} \langle \sum_i (x_i- x_0)^2 \rangle_t,
\end{eqnarray}
where $\langle.\rangle_s$ denotes a space average over the nodes of
the network, $\langle . \rangle_t$ denotes a time average, and $x_0$
is the fixed point of the uncoupled logistic map, $x_0=f(x_0)$.
$\sigma^2 =0$ if and only if $x_i=x_j$ $\forall$ $i$, $j$, while
$\sigma'^2 =0$ if and only if $x_i=x_0$ $\forall$ $i$. Thus,
$\sigma'^2$ allows to distinguish synchronization in the steady
state from synchronization in an time dependent state. In the former
case, both $\sigma^2$ and $\sigma'^{2}$ are zero, in the latter
case, only $\sigma^{2}=0$.

\section{Results}
\label{resultados}

In this section we present the results of the simulations. We
consider a network of $N=200$ logistic maps with $a=4$ interacting
with different topologies, as described above. The parameters of the
Gaussian delay distribution are $\tau_0=5$ and $c=2$. Similar
results are found for other values of $\tau_0$ and $c$, with $c$ large enough
\cite{nosotros_prl_2005}. To asses the role of distributed delays, we
compare with two limits: (i) instantaneous coupling, $\tau_{ij}=0$
$\forall$ $i$, $j$, and (ii) homogeneous delays, $\tau_{ij} =
\tau_0$ $\forall$ $i$, $j$ ($c=0$).

The simulations start from a random initial configuration, with
$x_i(0)$ randomly distributed in [0,1], and the maps evolve
initially without coupling, during a time interval $0< t<
max(\tau_{ij})$, because the integration of delayed equations
requires the knowledge of the past state of the system over a time
interval given by the maximum delay. After that, the coupling is
turned on. We neglect transient effects disregarding a few
thousands of iterations.

Networks of coupled elements usually show multistability:
different initial conditions lead to different final states, and
delayed coupling tends to increase the number of coexisting
states \cite{multistability,multistability2}. Multistability is also enhanced when the local dynamics of the
uncoupled maps presents two or more competing attractors
\cite{gallas_physicaA_2006}, which does not occur for the logistic
map with $a=4$. When the coupling is weak and the network is not
synchronized, we observe a large diversity of dynamical clustered states.
However, in this ''weak coupling regime''
the global synchronization indicators, $\sigma^2$ and $\sigma'^2$,
depend mainly on $\epsilon$ and only in a small extend
on the initial conditions, the distribution of
delays, the network topology and the connectivity (with the
exception of all-even delays, for them there is a synchronization
''island'' that will be discussed below). In the following, the
plots of $\sigma^2$ and $\sigma'^2$ are done by averaging over 10
different states generated from randomly chosen initial conditions. The only
exception is Fig. 6, that is done with just one
initial condition (that is the same for all $\epsilon$).

In the first section we characterize the network propensity for
synchronization in terms of the indicator $\sigma^2$; in the second
section, in terms of $\sigma'^2$. In these two sections the network
does not contain self-feedback links ($\eta_{ii}=0$ $\forall$ $i$).
In the last section we asses the role of feedback links, by studying
the same networks (with the same delay distributions and topologies)
but with a feedback link in each node ($\eta_{ii}=1$ $\forall$ $i$).

\subsection{Characterization in terms of $\sigma^2$}

\begin{figure}[tbp]
\resizebox{1.0\columnwidth}{!}{\includegraphics{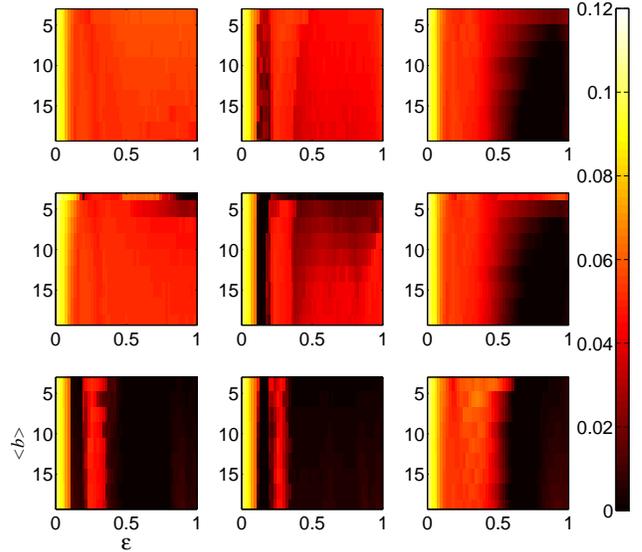}}
\caption{(Color online) Plot of $\sigma^2$ in the parameter space
(coupling strength, average number of neighbors) for three networks
with regular topologies: nearest-neighbors (NN), top row,
nearest-neighbors with central node (ST), middle row, and bottom row
{\it K-to-all} (KA). The delays are: zero (left column), homogeneous
($\tau_{ij}=5$ $\forall$ $i$ and $j$, central column), and
heterogeneous ( $\tau_0=5$, $c=2$, right column).}
\label{fig:RR-sin-auto}
\end{figure}

\begin{figure}[tbp]
\resizebox{1.0\columnwidth}{!}{\includegraphics{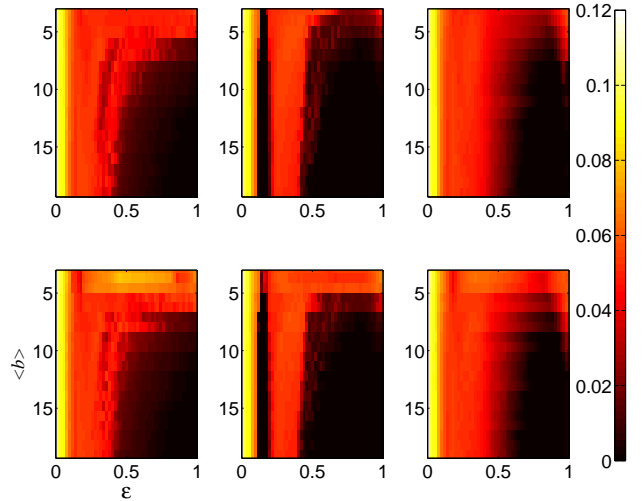}}
\caption{(Color online) Plot of the order parameter $\sigma^2$,
Eq.\ref{sp1} in the parameter space (coupling strength, average
number of neighbors) for two heterogeneous networks of 200 nodes: SW
(small-world, top row), SF (scale-free, bottom row). The delays are as
in Fig. \ref{fig:RR-sin-auto}: no-delays (left
column), homogeneous delays (center column), and heterogeneous
delays (right column). Other parameters as Fig. 1.} \label{fig:RI-sin-auto}
\end{figure}

Figures~\ref{fig:RR-sin-auto} and \ref{fig:RI-sin-auto} display
color-coded plots of the synchronization indicator $\sigma^2$ as a
function of the coupling strength, $\epsilon$, and the average
number of neighbors, $\langle b \rangle$, for the different
topologies and delay distributions.

Figure~\ref{fig:RR-sin-auto} displays results for the regular
topologies (NN: top row, ST: central row, KA: bottom row) and the
three distributions of delays considered (no-delays: left column,
homogeneous delays; central column, and distributed delays: right
column). Figure~\ref{fig:RI-sin-auto} displays results for the
heterogeneous networks (SW: top row, SF: bottom row) and the same
delay distributions.

We observe a general trend towards synchronization when increasing
the coupling strength and the average number of links. However we
note some important differences depending on the topology and on the
delays.

(i) {\it Heterogeneous delays}: if $\langle b \rangle$ is large
enough, the synchronizability does not dependent on the topology,
note the remarkable similarity of the five panels in the right
columns of Figs. \ref{fig:RR-sin-auto} and \ref{fig:RI-sin-auto},
for $\langle b \rangle \gtrsim 10$.

(ii) {\it No-delays and homogeneous delays} (left and central
columns in Figs.~\ref{fig:RR-sin-auto} and \ref{fig:RI-sin-auto}):
the topology makes a significant difference in regular networks
(Fig. \ref{fig:RR-sin-auto}), and is less important in heterogenous
networks (Fig. \ref{fig:RI-sin-auto}), although some differences can
be observed for small $\langle b \rangle$ ($\langle b
\rangle\lesssim 10$).

(iii) {\it Regular networks:} in Fig. \ref{fig:RR-sin-auto},
comparing the panels in the right and central columns we notice that
the network KA (bottom row) is the one with better propensity to
synchronization, while the network NN (top row) is the one
exhibiting poorer synchronizability. We also notice that ST (middle
row) has better synchronizability than NN when $\langle b \rangle$
is small, but there are no significant differences between them for large enough $\langle b \rangle$.

We note that for larger values of $\langle b \rangle$ (not shown in
Fig. \ref{fig:RR-sin-auto} because we focus on the weak connectivity
region) the NN network eventually synchronizes. Synchronization
occurs for $\epsilon$ above a certain value, $\epsilon^*$, that
decreases with increasing $\langle b \rangle$, in good agreement
with the results of Ref.\cite{kurths_physicaD_2005} (see in
particular Fig. 4(b) of \cite{kurths_physicaD_2005}, where the
variable $\alpha$ plays the role of $\langle b \rangle$ here;
increasing $\langle b \rangle$ resulting in a transition from local
to global coupling).

(iv) {\it Heterogeneous networks:} in Fig. \ref{fig:RI-sin-auto} we
do not observe a significant difference: the panels in the top and
bottom rows are similar, at least for $\langle b \rangle>5$.

(v) With homogeneous delays, middle column in Figs.
\ref{fig:RR-sin-auto} and \ref{fig:RI-sin-auto}, there is an
``island of synchronization'' in the region $\epsilon \sim$ 0.15 to
0.19. In spite of the fact that the coupling is weak, the network
tends to synchronize, regardless of the topology. This is due to the
fact that the delay times are even; for odd delays this
synchronization window does not exist (see, e.g., Fig. 1 in
\cite{atay_PRL_2004} and Fig. 2 in \cite{gallas_physicaA_2006}). In
the middle column of Fig. \ref{fig:RR-sin-auto} and
\ref{fig:RI-sin-auto} we observe that the ``synchronization island''
is robust and occurs for all the topologies.

\subsection{Characterization in terms of $\sigma'^2$}

\begin{figure}[tbp]
\resizebox{1.0\columnwidth}{!}{\includegraphics{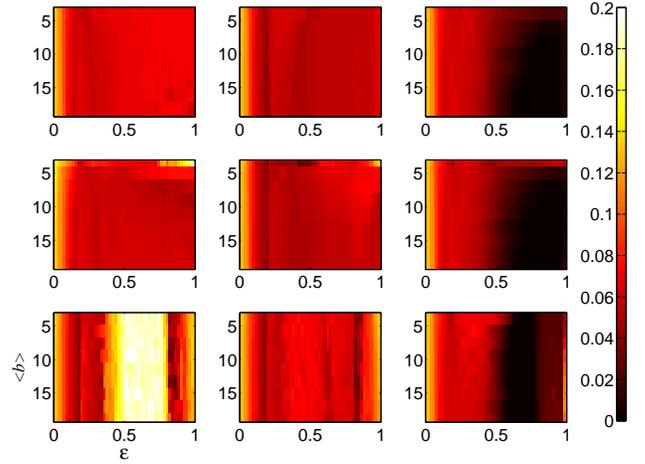}}
\caption{(Color online) Plot of the order parameter $\sigma'^2$, Eq.~(\ref{sp2}) 
in the parameter space (coupling strength, average number of neighbors) 
for three networks with regular topologies:
nearest-neighbors (NN), top row, nearest-neighbors with central node
(ST), middle row, and nearest-neighbors with two central nodes
(KA). Parameters are as in Fig. 1.} \label{fig:RR-sin-auto-sprima}
\end{figure}

\begin{figure}[tbp]
\resizebox{1.0\columnwidth}{!}{\includegraphics{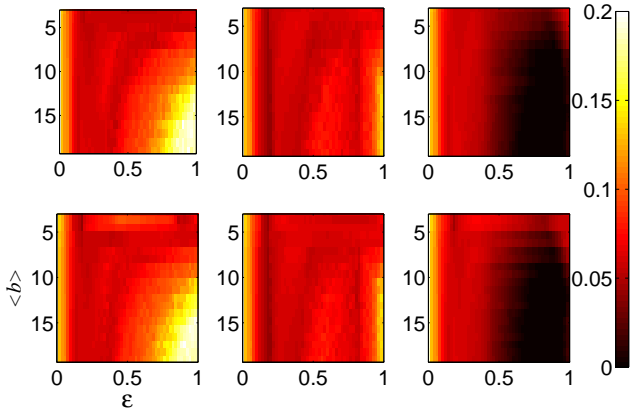}}
\caption{(Color online) Plot of $\sigma'^2$ in the parameter space (coupling strength, average number of neighbors)
 for three heterogeneous networks: SW (small-world, top row), RN (random
network, middle row), SF (scale-free, bottom row); and the different
delay distributions; instantaneous (left column), homogeneous delays
(center column), and heterogeneous delays (right column). Parameters are as in Fig.~1.}
\label{fig:RI-sin-auto-sprima}
\end{figure}

In order to investigate in more detail the two synchronization
regimes (time-dependent and fixed-point synchronization) now we
consider the synchronization indicator $\sigma'^2$, Eq. (\ref{sp2}),
that tends to zero only when the network synchronizes in the fixed
point. Figures~\ref{fig:RR-sin-auto-sprima} and
\ref{fig:RI-sin-auto-sprima} display color-coded plots of
$\sigma'^2$ as a function of the coupling strength, $\epsilon$, and
the average number of neighbors, $\langle b \rangle$, for the same
delay distributions and network topologies as Figs. 2 and 3
respectively.

In the case of heterogenous delays, again we notice a remarkable
similarity in the five panels in the right columns of Figs.
\ref{fig:RR-sin-auto-sprima} and \ref{fig:RI-sin-auto-sprima} for
$\langle b \rangle$ large enough, confirming that the
synchronizability of the network is largely independent of its
topology. However, we note that for the topology KA (bottom panel,
right column in Fig. \ref{fig:RR-sin-auto-sprima}) the network
losses synchrony if the coupling is too strong ($\epsilon
\thickapprox 1$).

 {\it Regular networks:} comparing the panels in the right and central columns of
Fig. \ref{fig:RR-sin-auto-sprima}, we can notice that the network KA
(bottom row) is the one with poorer synchronizability in the
steady-state.

{\it Heterogenous networks:} in Fig. \ref{fig:RI-sin-auto-sprima} is
observed that the SW and SF topologies have very similar propensity
for synchronization in the steady-state.

It can be noticed that, as the average number of links per node and
the coupling strength increase, the networks with heterogenous
delays tend to synchronize in the steady state, while the networks
with no-delays or with homogenous delays do not [in Figs.
\ref{fig:RR-sin-auto-sprima} and \ref{fig:RI-sin-auto-sprima},
compare the left and central columns with the right column]. This
tendency is particularly clear in heterogeneous networks, Fig.
\ref{fig:RI-sin-auto-sprima}, where in the left and in the central
columns $\sigma'^2$ is large for $\langle b \rangle \ge 15$ and
$\epsilon \ge 0.8$ (right, bottom corner of the panels), while in the
panels of the right column, $\sigma'^2$ is zero or very small in
that region.

The ``synchronization island'' discussed in the previous section is
not observed in Figs. \ref{fig:RR-sin-auto-sprima} and
\ref{fig:RI-sin-auto-sprima} because in this ''island'' the network
synchronizes in a time-dependent state that has $\sigma'^2$
different from zero.

\subsection{Connectivity threshold}

\begin{figure}[tbp]
\resizebox{1.0\columnwidth}{!}{\includegraphics{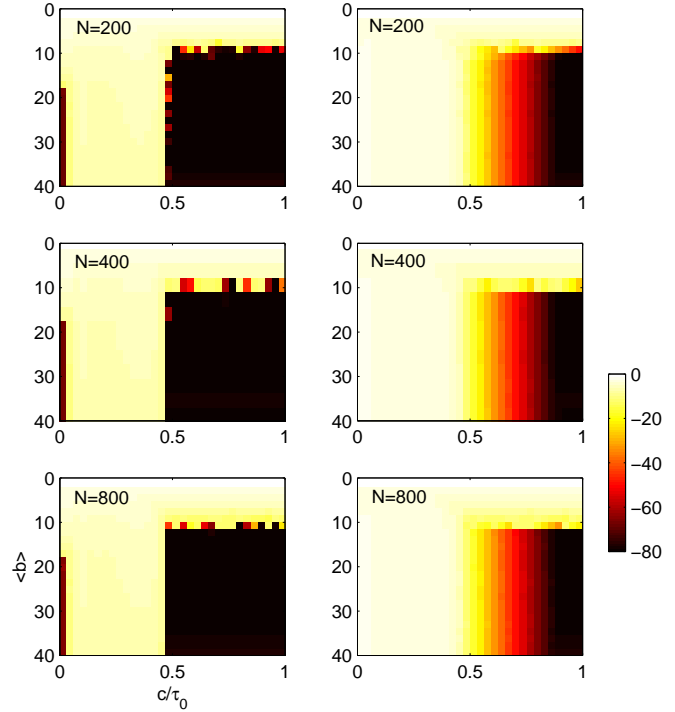}}
\caption{(Color online) Plot of $\sigma^2$ (left column) and $\sigma'^2$ (right column) in logarithmic scale, in the parameter space
(normalized width of the delay distribution, $c/\tau_0$, average connectivity, $\langle b \rangle$) for three
system sizes: $N=200$ (top row), $N=400$ (middle row), and $N=800$ (bottom row). Parameters are $\tau_0=5$ and $\epsilon=1$,  SW topology.}
\label{fig:RI-sin-auto-log}
\end{figure}

In the presence of heterogeneous delays, it can be expected that the average number
of neighbors has to be above a certain value for the distribution of delays to play an effective role. This is indeed shown in Fig.~\ref{fig:RI-sin-auto-log}, where we plot the synchronization indicators $\sigma^2$ and $\sigma'^2$ in the plane ($c/\tau_0$, $\langle b \rangle$) for various network sizes and the SW topology. $\sigma^2$ and $\sigma'^2$ are here plotted on a logarithmic scale to reveal the following features of the synchronization  transition: for connectivity below $\langle b \rangle \approx 10$, the network does not
synchronize, regardless of the network size and of the width of the delay distribution. Similar results are found for
other values of $\epsilon$ and $\tau_0$. For $\langle b \rangle \geq 10$, as $c$ increases and the delays become more heterogeneously distributed there is a sharp transition to isochronous synchronization, reveled in the plot of $\sigma^2$ (left column), which shows a sharp boundary between the dark and light regions, at $c/\tau_0$ slightly below $0.5$. The smooth transition seen in the plot of $\sigma'^2$ (right column) reveals that initially the network does not synchronize in the fixed point but in a time-dependent state, and as $c$ increases and the delay distribution enlarges, it gradually approaches the fixed point.

\subsection{Influence of self-feedback links}

We assess the effect of self-feedback links by setting $\eta_{ii}=1$
$\forall$ $i$, in the previously studied network topologies.

\begin{figure}[tbp]
\resizebox{1.0\columnwidth}{!}{\includegraphics{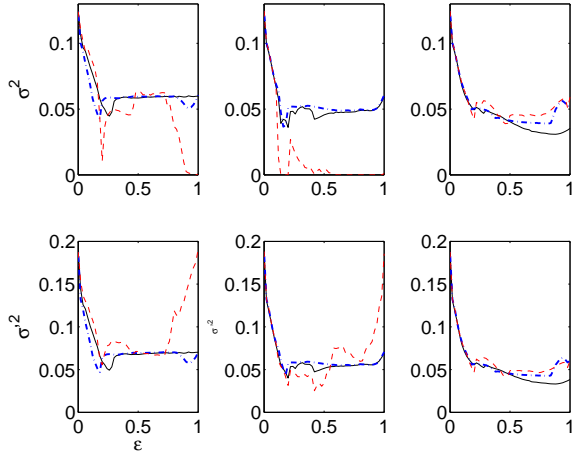}}
\caption{(Color online) Plot of $\sigma^2$ (top row) and
$\sigma^{'2}$ (bottom row) vs. the coupling strength for a NN
network with two neighbors per node and a feedback link (solid
line), for the same ring without feedback links (dot-dashed
line, blue online) and for the ST network composed by the ring plus a central
node (dashed line, red online). The delays and other parameters are as in the previous figures:
no-delays (left column), homogeneous delays (central column) and
heterogeneous delays (right column).}
\label{fig:ef_autolink-NNvsST}
\end{figure}

In a large enough network (as the one studied so far, with $N=200$
maps), the simulations show that feedback links have a small
influence on the network synchronizability when the connectivity is
low (when $\langle b \rangle\lesssim 0.5 N$), and their effect is
negligible when $\langle b \rangle$ is larger. With 2D color-coded
plots is difficult to appreciate the influence of the feedback
links; therefore, we plot in Fig. ~\ref{fig:ef_autolink-NNvsST} the
global synchronization indicators vs. the coupling strength, for a
fixed value of $\langle b \rangle$, that is as low as possible.

We consider three regular topologies: (i) a ring with two neighbors
per node and a self-feedback link in each node (black solid line);
(ii) the same ring without the self-feedback links (dot-dashed line,
blue online); and (iii) a ST network composed by the ring with two
neighbors per node (without self-feedback links) and a
centrally connected node (dashed line, red online). It can be
noticed that with no delays and with uniform delays (left and
central columns in Fig. ~\ref{fig:ef_autolink-NNvsST}), the ST
network synchronizes for $\epsilon$ large enough, while the ring,
with and without self-feedback links, does not. With randomly
distributed delays (right column in Fig.
~\ref{fig:ef_autolink-NNvsST}), the three networks do not
synchronize for any value of $\epsilon$.

Let us now consider the influence of feedback links in a smaller
network of $N=20$ maps. Figure 6 displays the global synchronization
indicator while Figs. 7-9 present some examples of the space and
time evolution of the network. The figures are done by representing
in a color scale the variable $x_i(t)$, with the space index $i$ on
the horizontal axis and the time index $t$ on the vertical axis. The
left column of Figs. 6-9 shows results for networks without feedback
loops, and the right column, for the same networks with feedback loops. A
relevant effect of the feedback can be observed on both, the global
macroscopic indicator and on the microscopic network configuration.
There are situations in which feedback loops enhance coherence,
giving rise to spatially more ordered patterns (e.g., in Fig. 7, top
and bottom row), while in others, on the contrary, feedback loops
destroy the spatial coherence of the pattern (e.g., in Fig. 9).
Interestingly, it can be observed that with random delays, Fig. 9,
the synchronization is not always on a homogeneous state, but there
are also static patterns with spatial ''antiphase'' arragement. The
characterization of these patterns is in progress and will be
reported elsewhere.

\begin{figure}[tbp]
\resizebox{1.0\columnwidth}{!}{\includegraphics{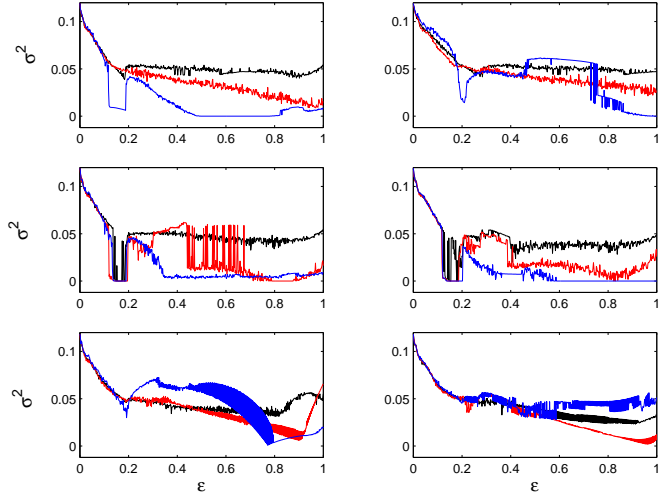}}
\caption{(Color online) Synchronization indicator $\sigma^2$ for a
small network of $N=21$ nodes without feedback loops (left
column), and with feedback loops (right column).  The delay
distributions are: no delays (top row), fixed delays (middle row),
and distributed delays (bottom row). The topologies are: a
nearest-neighbors ring (black), a ring with the addition of a
central node (red) and a pure star network composed by a central hub
connected to $N=20$ nodes (blue).}
\end{figure}

\begin{figure}[tbp]
\resizebox{1.0\columnwidth}{!}{\includegraphics{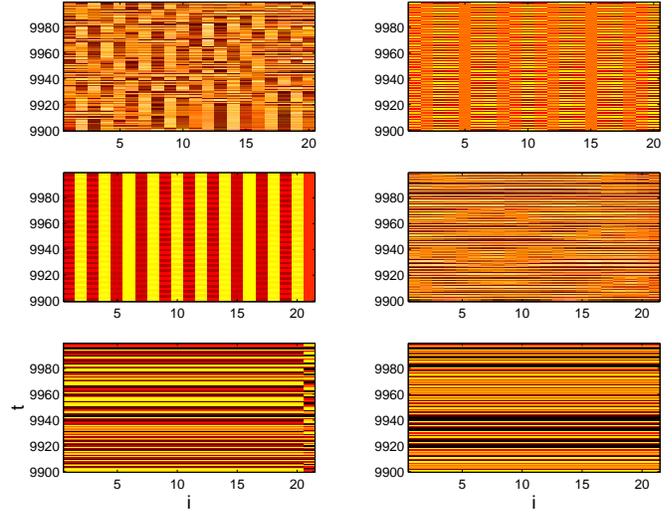}}
\caption{(Color online) Synchronization patterns with instantaneous
interactions. The network topologies are: a nearest-neighbors ring
of $N=20$ nodes (top row); a $N=20$ nearest-neighbors ring with the
addition of a central node (middle row) and a star network composed
by a central hub connected to $N=20$ nodes (bottom row). Without feedback
loops (left column), with feedback loops (right column). The
coupling strength is $\epsilon=1$}
\end{figure}

\begin{figure}[tbp]
\resizebox{1.0\columnwidth}{!}{\includegraphics{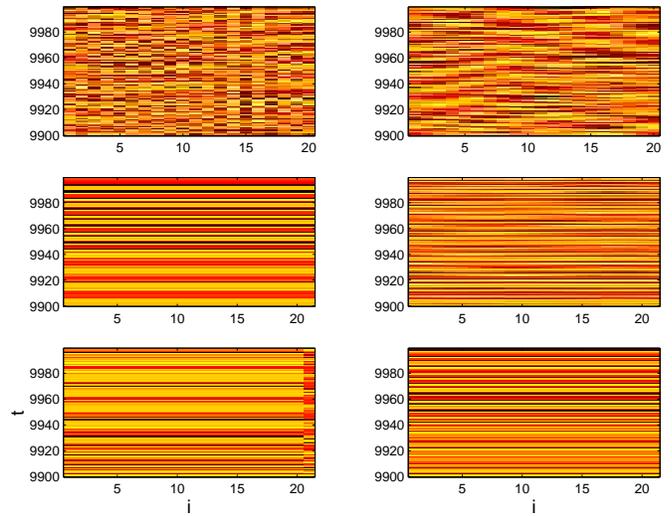}}
\caption{(Color online)  Synchronization patterns with homogeneous
delays ($\tau_0=5$, $c=0$). The network topologies are as in Fig. 7.
(a),(b) $\epsilon=1$; (c)-(f) $\epsilon=0.8$. Without feedback
loops (left column), with feedback loops (right column).}
\end{figure}

\begin{figure}[tbp]
\resizebox{1.0\columnwidth}{!}{\includegraphics{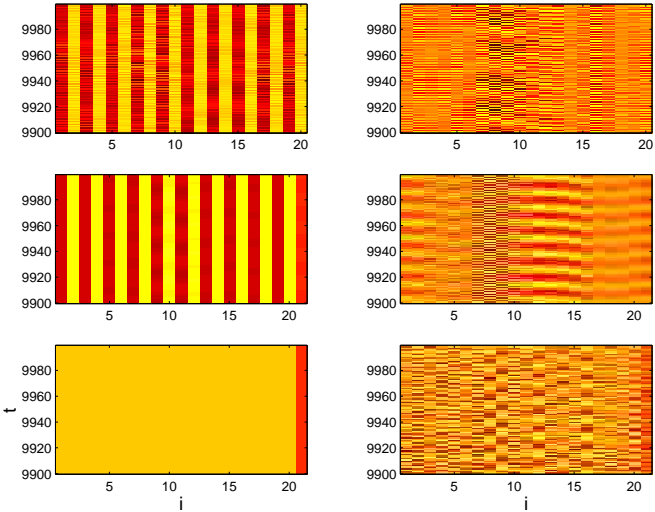}}
\caption{(Color online) Synchronization patterns with heterogeneous
delays ($\tau_0=5$, $c=2$). The network topologies are as in Figs. 7
and 8. (a),(b) $\epsilon=0.9$; (c),(d) $\epsilon=1$; (e),(f)
$\epsilon=0.8$. Without self-feedback
loops (left column), with self-feedback loops (right column).}
\end{figure}

It is also interesting to notice that when the coupling strength is
small (roughly speaking, when $\epsilon < 0.2$) the dependence of
the global synchronization indicators $\sigma^2$ and $\sigma'^{2}$
with $\epsilon$, shown in Figs. ~\ref{fig:ef_autolink-NNvsST} and 6,
is very similar. This suggests that the global dynamics is almost
independent of the topology, the connectivity, the network size and
the delay distribution. We refer to this region as the ''weak
coupling'' region. It has been recently shown that in this region
all the nodes exhibit a qualitatively similar symbolic dynamics,
that, for instantaneous interactions, depends mainly on the network
architecture and only to a small extent, on the local dynamics
\cite{atay_chaos_2006}.

\section{Summary and Conclusion}
\label{summary} \label{conclusions}

We studied the synchronizability of a network focusing on the roles
of the connectivity, the topology, and the delay times that are
associated with the links. The nodes were modeled by chaotic
logistic maps, and various topologies and delay distributions were
considered. For low connectivity (roughly speaking, when the mean
number of links per node is $\langle b \rangle < 0.1N$), and for
weak coupling ($\epsilon <0.1$), the network displays an irregular
oscillatory dynamics, regardless of the network topology and the
delay distribution. For large enough connectivity and coupling, the
synchronization is mainly determined by the delay
times: when the delays are homogeneously distributed the network
shows collective synchronous time-dependent oscillations; when
the delays are sufficiently distributed, the network synchronizes in
a spatially homogeneous steady-state. The propensity towards these
two synchronization regimes was characterized in terms of two
indicators, $\sigma^2$ and $\sigma'^2$ [Eqs. (\ref{sp1}) and
(\ref{sp2})], the first one tends to zero when the network
synchronizes isochronously [$x_i(t)=x_j(t)$ $\forall$ $i$, $j$],
while the second one tends to zero only when the network
synchronizes in the steady state [$x_i(t)=x_0$ $\forall$ $i$, with
$x_0=f(x_0)$ being the fixed point of the uncoupled maps].

When the coupling strength is weak the dependence of
the synchronization indicators with $\epsilon$ is very similar for
all topologies and delay distributions considered. This suggest
that in this region of ''weak coupling'' the network topology and
the delay distribution play no relevant role in the dynamics.
However, is important to remark that the global synchronization
indicators employed, $\sigma^2$ and $\sigma'^2$, have the limitation
that they characterize only isochronal synchronization and
steady-state synchronization respectively. In a complex network
where the interactions among the units are not instantaneous, other
synchronization patterns are also expected, such as states where the
nodes are synchronized but with lag-times between them
\cite{zanette_2003,lasers_2007}. An interesting indicator to analyze
in future studies is the one that measures the average distance
between the present state of a map and the delayed state of the maps
interacting with it: $$\sigma^2 = (1/N) \sum_{i}(1/b_i) \sum_j
\eta_{ij} \langle(x_i(t)- x_j(t-\tau_{ij}))^2 \rangle_t.$$ Also, it
will be very interesting to analyze synchronization patterns
using symbolic dynamics \cite{atay_chaos_2006} and complexity
indicators \cite{complexity,rosso}.

{\it With heterogeneous delays we also found that there is a connectivity threshold 
below which the network does not synchronize, regardless of the network size.
This minimum average number of neighbors is also independent of the delay distribution. 
Above the minimum connectivity level, as the width of the distribution increases, the plot of  $\sigma^2$ reveals that there is a sharp transition to isochronous synchronization, while the plot of  $\sigma'^2$ reveals that the network does not synchronize in the fixed point but in a time-dependent state, that gradually approaches the fixed point as the delays become more heterogeneous. Simulations were preformed for SW topology; the study of how this connectivity  threshold depends on the network topology is in progress and will be reported elsewhere.}

We studied the influence of feedback loops in each node
of the network; these feedback loops having the same delay distribution as the mutual
interactions, and found that when the network is large, feedback
loops have very little impact on the global synchronization indicators.
However, they affect synchronizability of small
networks, enhancing or degrading the synchronization, depending on
the network architecture and on the delay distribution. As a future
study it will be interesting to analyze the interplay of delayed
feedback with instantaneous mutual coupling and vice-versa, instantaneous feedback with delayed coupling.

\section{Acknowledgments}
ACM and MPC acknowledge financial support from CSIC and  PE\-DE\-CI\-BA (Uruguay). CM acknowledges support from the \textquotedblleft Ramon
y Cajal\textquotedblright\ Program (Spain) and the European
Commission (GABA project, FP6-NEST 043309).

\end{document}